\documentclass[
reprint,
superscriptaddress,
nofootinbib,
amsmath,amssymb,
aps,
prl,
]{revtex4-2}

\usepackage{graphicx} 
\usepackage{dcolumn} 
\usepackage{bm} 
\usepackage[colorlinks=true,allcolors=blue]{hyperref} 

\usepackage[version=4]{mhchem}

\begin{document}
\setcounter{secnumdepth}{3}

\preprint{APS/123-QED}

\title{The Belousov-Zhabotinsky reaction reveals two regimes of non-Arrhenius temperature scaling in relaxation oscillators.
}

\author{Simen Jacobs}
\affiliation{Laboratory of Dynamics in Biological Systems, Department of Cellular and Molecular Medicine, KU Leuven, Herestraat, 49, Leuven, Belgium}
\author{Nikita Frolov}
\affiliation{Laboratory of Dynamics in Biological Systems, Department of Cellular and Molecular Medicine, KU Leuven, Herestraat, 49, Leuven, Belgium}
\author{Panna Farkas}
\affiliation{Hevesy György Doctoral School of Chemistry, Eötvös Loránd University, Pázmány Péter sétány 1/A, Budapest, Hungary.}
\author{István Lagzi}
\affiliation{Department of Physics, Institute of Physics, Budapest University of Technology and Economics, Muegyetem rkp 3., H-1111 Budapest, Hungary}
\affiliation{HUN-REN–BME, Condensed Matter Physics Research Group, Budapest University of Technology and Economics, Muegyetem rkp 3., H-1111 Budapest, Hungary}
\author{István Szalai}
\affiliation{Laboratory of Adaptive and Autonomic Materials, Institute of Chemistry, Eötvös Loránd University, Pázmány Péter sétány 1/A, Budapest, Hungary.}
\author{Lendert Gelens}
\affiliation{Laboratory of Dynamics in Biological Systems, Department of Cellular and Molecular Medicine, KU Leuven, Herestraat, 49, Leuven, Belgium}

\date{\today}

\begin{abstract}
    The period of biological and chemical oscillators scales with temperature in a characteristic way. Some oscillators are very well described by an Arrhenius law, while others show systematic deviations. Several frameworks have been proposed to explain such deviations, but they are either phenomenological, focus on activation energy imbalances in specific circuits, or restrict themselves to sequential processes. Here we develop a mechanistic account of the temperature scaling of relaxation oscillators, using the Belousov-Zhabotinsky (BZ) reaction as a model system. 
    We distinguish two typical scenarios by their temperature-scaling signatures.    
    In the first, an Arrhenius-dependent timescale separation parameter produces a biphasic Arrhenius scaling of the period as the oscillator approaches a Hopf bifurcation. In the second, Arrhenius-dependent nullclines hide the same bifurcation behind a canard explosion, yielding apparent single-line Arrhenius scaling.
    Measuring the electrode potential of a classical and an uncatalyzed BZ reaction, over a very wide temperature range ($\approx$ 100 °C), and comparing to dynamical models, we find that the two reactions represent these two distinct scenarios. Furthermore, we show that a single parameter characterizing the waveform asymmetry between fast and slow phases quantitatively predicts the temperature scaling of three other observables close to the Hopf bifurcation: the period, amplitude, and phase noise.
    This analysis also recovers elementary activation energies of the BZ mechanism, including a new estimate for the autocatalytic step. We discuss how this framework and its waveform-based diagnostics apply to the analysis of general biochemical relaxation oscillators. 
\end{abstract}


\maketitle

\begin{figure*}[t]
    \centering
    \includegraphics[width=1\linewidth]{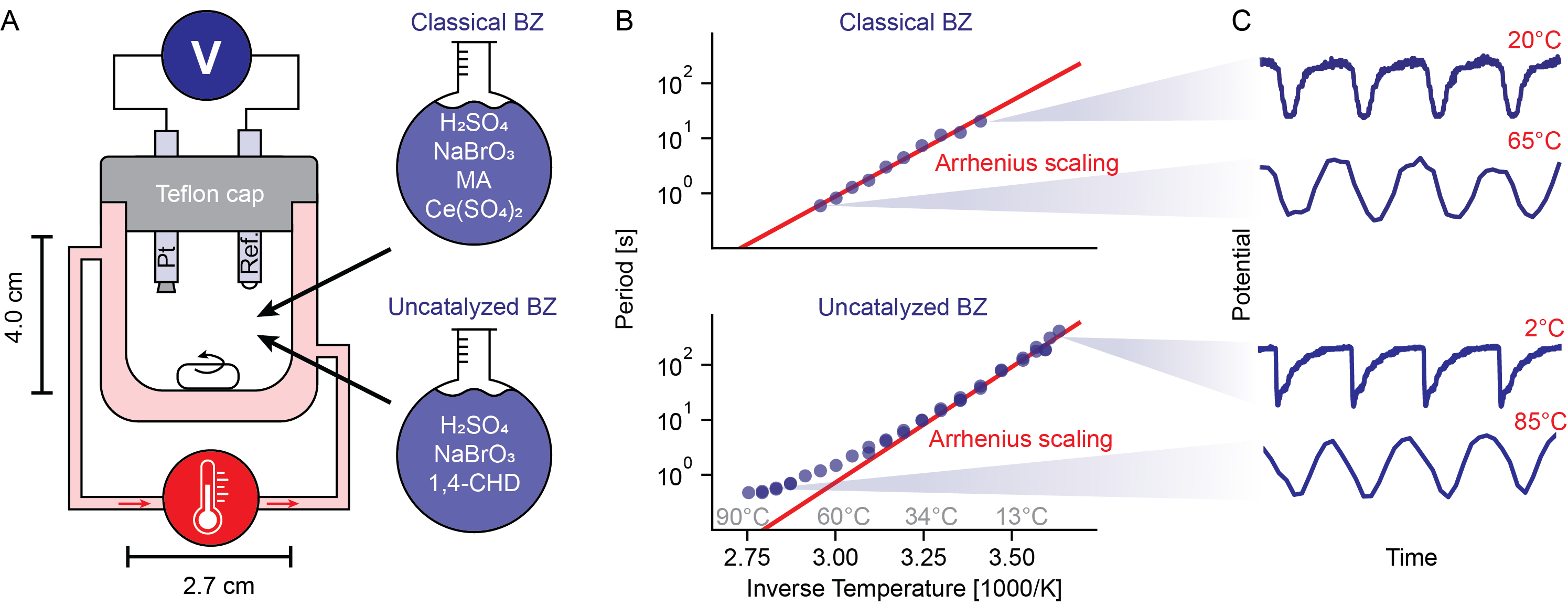}
    \caption{\textbf{The temperature scaling of the classical and uncatalyzed BZ reaction.} A: Sketch of the experimental setup showing the dimensions of the reaction vessel with magnetic stirrer at the bottom, a thermostat to regulate the temperature, and a Pt/Reference electrode pair on top to record the signal. The necessary components to start each of the BZ reactions are also shown (MA: malonic acid; 1,4-CHD: 1,4-cyclohexadione). For more, see Section I of the Supplemental Material. B: Temperature scaling of the period of oscillations for the two studied reactions in an Arrhenius diagram (x-axis: $1/T$; y-axis: period on a logarithmic scale). For reference, we included straight lines to represent ideal Arrhenius scaling. C: Example of the measured electrode signal at selected temperatures. Both the period and the shape of the oscillations are temperature-dependent. For each of the plotted signals, the x-axis and y-axis have been rescaled separately to facilitate the comparison of waveforms.}
    \label{fig:introduction}
\end{figure*}

\section{Introduction}

Across scales from biomolecules to ecosystems, biological rates depend on temperature in ways that often deviate from the simple Arrhenius behavior expected for elementary chemical reactions \cite{Dell2011, brown_toward_2004, kontopoulos_no_2024, arnoldi_universal_2025, jacobs_understanding_2026, jacobs_understanding_2026-1}. This is especially true for biological oscillators, where the period as a function of temperature frequently shows a characteristic curvature in the Arrhenius plot rather than the straight line that single-step kinetics would predict \cite{pittendrigh1954temperature, hastings1957mechanism, rombouts_mechanistic_2025, fu_temperature_2024}. Several mechanistic frameworks have been proposed to account for such non-Arrhenius scaling, including the averaging of activation energies across cascades of many elementary reactions \cite{voits_generic_2025, jacobs_beyond_2026}, and the imbalance of activation energies between opposing reactions in specific oscillatory circuits such as the early embryonic cell cycle \cite{rombouts_mechanistic_2025} and circadian clocks \cite{ruoff1995antagonistic, fu_temperature_2024}. Yet these explanations are circuit-specific: each identifies a particular biochemical asymmetry within a particular oscillator. It is not known whether non-Arrhenius scaling has a deeper origin in the shared dynamical structure of \textit{relaxation oscillators} — a class that encompasses many of the most-studied biological oscillators, including the embryonic cell cycle and circadian rhythms \cite{rombouts_mechanistic_2025, fu_temperature_2024}.

Relaxation oscillators are nonlinear oscillators characterized by the presence of well-separated fast and slow timescales within one cycle \cite{ginoux_van_2012}. Because their dynamics emerge from the coordination of these timescales, the temperature response of a relaxation oscillator is highly sensitive to which reactions set the slow manifold. If the rate-limiting slow process is temperature-insensitive, or if temperature-sensitive rates are arranged to cancel along this manifold, the oscillator can be remarkably robust to temperature changes — as observed in temperature-compensated circadian clocks \cite{pittendrigh1954temperature, hastings1957mechanism, fu_temperature_2024}. Conversely, if multiple temperature-dependent rates shift the relative position of fast and slow processes, both the period and the very existence of oscillations can be altered. What remains unclear is how these different regimes arise from the underlying dynamical structure of a relaxation oscillator, and how the transitions between them can be probed experimentally across a wide temperature range.

Here, we combine experimental measurements with theory and modeling to address this gap, using the Belousov--Zhabotinsky (BZ) reaction \cite{zhabotinsky1964periodic, zaikin1970concentration, field1972oscillations} as our model system. The experiments resolve oscillations with fine temperature spacing and high temporal resolution, allowing us to extract not only the period but also the detailed waveform shape at each temperature. As a paradigmatic chemical relaxation oscillator, the BZ reaction is well-suited to this purpose: its mechanism is well-characterized, reduced theoretical models exist, and reaction conditions can be tightly controlled. We performed measurements in two versions of the reaction: the classical catalyzed variant, in which \ce{Ce^4+} acts as the catalyst, and an uncatalyzed variant \cite{koros1978uncatalysed, szalai1998} in which an organic compound replaces the reduced form of the catalyst in the autocatalytic pathway (see Fig. \ref{fig:introduction}A). A key practical advantage of the uncatalyzed variant is that \ce{CO_2} formation does not disturb the observations, allowing us to measure sustained oscillations over an exceptionally wide temperature range of approximately 100 °C. All details about the experimental procedures are given in Section I of the Supplemental Material.

A simple approximation for the temperature dependence of chemical oscillators was proposed by Körös, who suggested an Arrhenius-type relationship between the period ($P$) and absolute temperature ($T$):
\begin{align} \label{eq:Arrhenius}
    \log P = \frac{E}{RT} + \log C,
\end{align}
where $E$ is an effective activation energy, $C$ a temperature-independent constant, and $R$ the ideal gas constant \cite{koros1974monomolecular}. This relationship may seem surprising, since the Arrhenius law was originally formulated for the rate constants of elementary chemical reactions, whereas oscillator periods emerge from a network of many such reactions. Nevertheless, analogous Arrhenius-type fits have been independently applied to the period of biological relaxation oscillators \cite{dolbear_cricket_1897,mair_control_2005,Dell2011,begasse_temperature_2015,mundim_temperature_2020,arnoldi_universal_2025,rombouts_mechanistic_2025}, and the resulting effective activation energy (often expressed as a $Q_{10}$-value) has been widely used to characterize the temperature response of the oscillators.

\begin{figure*}[t!]
    \centering
    \includegraphics[width=1\linewidth]{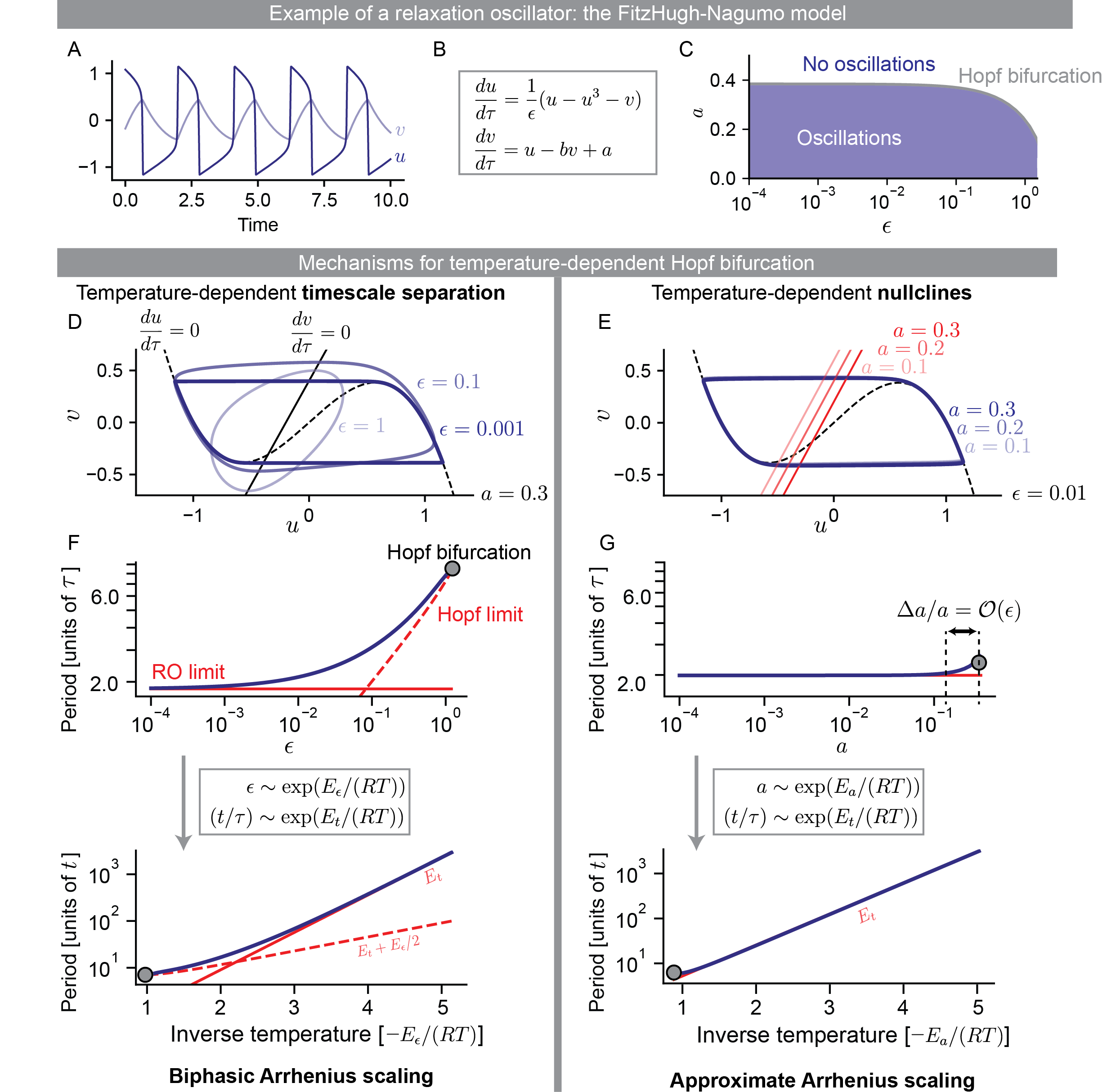}
    \caption{\textbf{The FitzHugh--Nagumo (FHN) model with Arrhenius-dependent parameters.} Analytical derivations and simulation parameters can be found in Section III and VIII of the Supplemental Material. 
    A: Typical timeseries of the FHN relaxation oscillator. The timescale separation is clearly visible in the rapid switches between the rising and falling of $v$. 
    B: Defining equations of the FHN model. 
    C: Parameter space, showing the oscillatory region in the $(a,\epsilon)$ plane for $b = 0.5$.  
    D: Nullclines of the FHN model for varying values of $\epsilon$ and $b = 0.5, a = 0.3$. 
    E: Nullclines of the FHN model for varying values of $a$ and $b = 0.5, \epsilon = 0.01$. 
    F: Scaling of the period of the FHN model with $\epsilon$ and with temperature, assuming an Arrhenius law for $\epsilon$ (and all other parameters as in D). This results in a biphasic Arrhenius scaling, for which the Hopf limit (dashed red line and black point) and relaxation oscillator limit (red line) can be analytically derived. 
    G: Scaling of the period of the FHN model with $a$ and with temperature, assuming an Arrhenius law for $a$ (and all other parameters as in E). This results in an approximate Arrhenius law over almost the entire temperature range. The relaxation oscillator limit (red line) and Hopf bifurcation point (black) can be analytically derived. }
    \label{fig:FHN}
\end{figure*}

Fig.~\ref{fig:introduction}B shows the measured periods of the classical and uncatalyzed BZ reactions in such an Arrhenius diagram. While the classical reaction is well approximated by a single Arrhenius line, the uncatalyzed reaction displays a clear convex curvature. Our analysis will show that this difference is not an artifact of the measurement range, but reflects a fundamental difference in how the internal timescales of the two reactions depend on temperature. That the timescale separation is itself temperature-dependent is directly visible in the waveform shape (Fig.~\ref{fig:introduction}C), which evolves from sawtooth-like oscillations at low temperatures, with a slow rising and fast falling potential, to more sinusoidal oscillations at high temperatures.

To explain these observations and to understand how the two regimes arise from the fast-slow structure of the oscillator, we combine experiments on the BZ reaction with a general dynamical framework. We first show that in a prototypical relaxation oscillator, a temperature-dependent timescale separation produces a characteristic biphasic Arrhenius scaling of the period as the system approaches a supercritical Hopf bifurcation, distinct from the canard-explosion scenario that arises when temperature affects only the nullclines. We then construct reduced temperature-dependent models of the classical and uncatalyzed BZ reactions and confront these models with the experimental data. A single experimental observable -- the waveform asymmetry -- quantitatively predicts the scaling of three others close to the Hopf bifurcation: the oscillation period, amplitude, and phase noise. The classical BZ reaction falls into the canard-explosion regime, while the uncatalyzed reaction enters the biphasic regime as its timescale separation is lost at high temperatures, providing direct experimental evidence for both regimes within a single chemical system. These results suggest that the two regimes -- and the crossover between them -- provide a unifying framework for non-Arrhenius temperature scaling in chemical and biological relaxation oscillators.

\section{Temperature-dependent timescale separation produces biphasic Arrhenius scaling} \label{sec:theory}
When a collection of (bio)chemical reactions is reduced to a low-dimensional system of ordinary differential equations (ODEs), the parameters in the resulting model typically arise as products and quotients of individual rate constants. (In Sections IV and V of the Supplemental Material we show how this works for both BZ reactions.) Their temperature dependence is therefore approximately Arrhenius-like, with effective activation energies that can be positive or negative \cite{ruoff1995antagonistic, pullela_temperature_2009}. As the temperature is varied, these parameters change, and the dynamical behavior of the oscillator can undergo bifurcations.

To make this concrete, we work with a general two-dimensional relaxation oscillator for variables $u, v$ with a timescale separation parameter $\epsilon$:
\begin{align} \label{eq:rel_osc}
    du/d\tau &= \frac{1}{\epsilon} F(u,v;\bm{a}), \nonumber \\
    dv/d\tau &= G(u,v;\bm{a}),
\end{align}
where $\bm{a}$ collects the remaining parameters. We assume that (i) $\epsilon \ll 1$ and (ii) $F/G = \mathcal{O}(1)$, so that $u$ evolves on a fast timescale and $v$ on a slow one. Writing the system in this form may require rescaling the physical time $t$ by a constant factor $(t/\tau)$. Throughout this section, we illustrate the analysis using the FitzHugh--Nagumo (FHN) model (Fig. \ref{fig:FHN}A,B), a prototypical relaxation oscillator \cite{fitzhugh_impulses_1961,nagumo_active_1962,cebrian-lacasa_six_2024}. The necessary mathematical background to derive the statements in the rest of this section are given in Section II of the Supplemental Material, while Section III of the Supplemental Material describes the application to the FHN oscillator.

The geometry of the limit cycle is most easily understood through the nullclines, the curves where $du/d\tau = 0$ and $dv/d\tau = 0$ (Fig. \ref{fig:FHN}D,E). For small $\epsilon$, the limit cycle closely follows the nullcline of the fast $u$-variable for most of its trajectory, and a quasi-steady-state approximation gives, to lowest order, a period that is independent of $\epsilon$ (see Supplemental Material, Section II A). We refer to it as the \textit{RO limit} (for relaxation oscillator limit):
\begin{align} \label{eq:relosc_period_tau}
    P_\tau = P_{ro} + \mathcal{O}(\epsilon).
\end{align}

Limit cycle oscillations can be created or destroyed by a number of mechanisms — supercritical and subcritical Hopf bifurcations, saddle-node bifurcations of cycles, SNIC bifurcations, and so on — but the supercritical Hopf bifurcation is one of the most common and is also the one we will encounter in both the classical and uncatalyzed BZ reaction. Close to such a bifurcation, the timescale separation breaks down: assumption (i) or (ii) imposed earlier on the relaxation oscillator fails, and the RO limit \eqref{eq:relosc_period_tau} no longer applies. A linear stability analysis instead gives a different limiting expression, which we refer to as the \textit{Hopf limit} (see Supplemental Material, Section II B):
\begin{align} \label{eq:hopf_period_tau}
    P_\tau = P_{h} = A\sqrt{\epsilon} + \mathcal{O}(\epsilon),
\end{align}
where $A$ is a constant independent of $\epsilon$. The locus of the supercritical Hopf bifurcation in the $(\epsilon, a)$ parameter space for the FHN model is shown in Fig. \ref{fig:FHN}C: varying $\epsilon$ (failure of (i)) or varying $\mathbf{a}$ (failure of (ii)) can each drive the system from the RO limit toward the Hopf limit.\\

\textbf{Two ways for temperature to drive a Hopf bifurcation.}
With the RO and Hopf limits in hand, we can now ask how an Arrhenius dependence of the underlying parameters translates into a temperature dependence of the period. Two scenarios emerge, based on which parameter carries the temperature dependence.\\

\textit{Case 1: Arrhenius-dependent timescale separation.}\\
Suppose that
\begin{align}
    \log \epsilon \sim \frac{E_\epsilon}{RT}, \qquad \text{and} \qquad \log (t/\tau) \sim \frac{E_t}{RT},
\end{align}
while the remaining parameters are temperature-independent (or depend only weakly on temperature). Without loss of generality and in agreement with the trends in Fig. \ref{fig:introduction} we take $E_\epsilon < 0$. The nullclines of \eqref{eq:rel_osc} are then temperature-invariant, but the timescale separation between fast and slow variables decreases as temperature increases. The full period $P = (t/\tau) P_\tau$ scales \textit{biphasically}: at low temperatures where $\epsilon$ is small, the system sits near the RO limit and
\begin{align} \label{eq:ROperiod}
    \log P \sim \frac{E_{t}}{RT},
\end{align}
while at higher temperatures, as $\epsilon$ grows and the system approaches the Hopf limit,
\begin{align}
    \log P \sim \frac{E_{t}+E_{\epsilon}/2}{RT}.
\end{align}
In the Arrhenius diagram, this scaling appears as two straight lines with different slopes joined by a transition region (Fig. \ref{fig:FHN}F). The crossover from the RO limit to the Hopf limit is the signature of an oscillator approaching a Hopf bifurcation through a loss of timescale separation.\\

\textit{Case 2: Arrhenius-dependent nullclines.}\\
Suppose instead that $\epsilon$ is (effectively) temperature-independent, and that the Hopf bifurcation is driven by an Arrhenius dependence of one of the system parameters $\bm{a}$. The nullclines now shift with temperature, but the timescale separation remains intact across the entire oscillatory range. The consequence is dramatic: because $\epsilon$ stays small, the transition from the RO limit to the Hopf limit happens over a parameter change of order $\Delta a = \mathcal{O}(\epsilon)$ — a phenomenon known as a \textit{canard explosion} \cite{krupa_relaxation_2001, desroches_mixed-mode_2012, kuehn_multiple_2015} (see Supplemental Material, Section II C). In the Arrhenius diagram, the Hopf limit occupies such a narrow temperature window that it is typically invisible in experiment: the period appears to follow the single Arrhenius law of \eqref{eq:ROperiod} across the entire accessible temperature range, even though the system is in fact approaching a bifurcation (Fig. \ref{fig:FHN}G).

These two cases possess qualitatively different fingerprints in the Arrhenius diagram: biphasic curvature versus apparent Arrhenius scaling. Both arise from the same underlying mechanism, through a temperature-induced supercritical Hopf bifurcation in a relaxation oscillator. In the next section, we show that the classical and uncatalyzed BZ reactions occupy these two regimes respectively, and that the experimental data bear the predicted fingerprints.

\begin{figure*}[t!]
    \centering
    \includegraphics[width=1\linewidth]{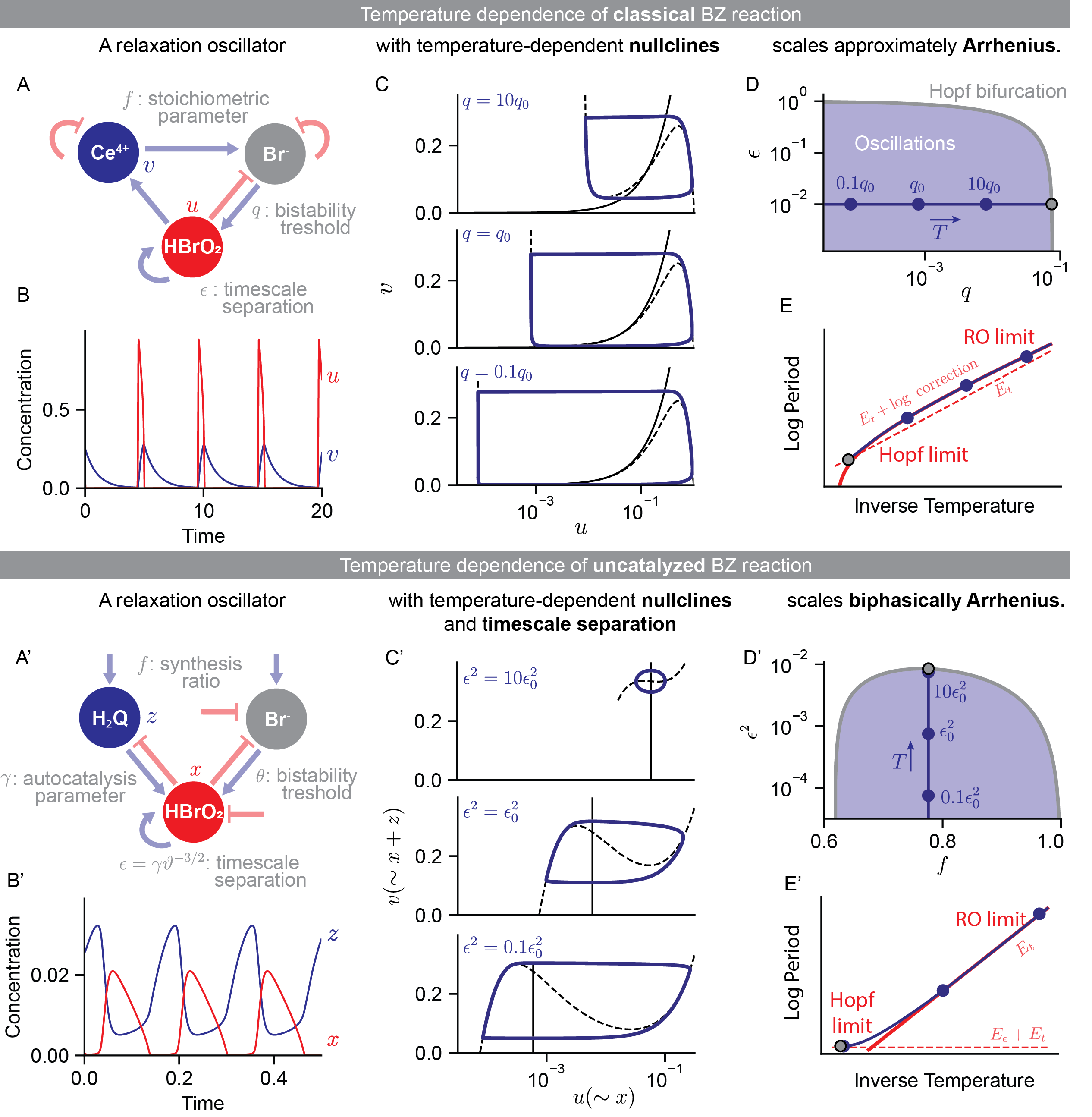}
    \caption{\textbf{Temperature-dependent models of oscillations in the classical and uncatalyzed BZ reaction.} Analytical derivations are given in Sections IV and V of the Supplemental Material, details on the numerical simulations and parameter values in Section VIII of the Supplemental Material. 
    A/A': Interactions between the oscillating intermediates in the chemical reaction. In both reactions \ce{Br-} (in grey) is eventually removed from the dynamical equations through a quasi-steady state approximation. 
    B/B': Timeseries for a typical set of parameters at room temperature. 
    C/C': Phase plane plot showing a limit cycle and the $u/v$-nullcline where $du/dt = 0$ or $dv/dt = 0$ for different values of a control parameter. The parameter values $q_0 = 8 * 10^{-4} $ and $\epsilon_0 = 7.5 * 10^{-4} $ are typical at room temperature (see Supplemental Material, Section VIII). 
    D/D': Parameter space, showing the oscillatory region and the locus of the Hopf bifurcation. If the control parameters $q$ or $\epsilon$ are Arrhenius-dependent (see \eqref{eq:temp_epsilon} and \eqref{eq:temp_q}) with negative activation energies the system undergoes a Hopf bifurcation at high temperatures. 
    E/E': Characteristic temperature scaling of the period of both systems in an Arrhenius diagram. The inverse temperature axis has arbitrary units since it can be rescaled by adjusting activation energies and pre-exponential factors. }
    \label{fig:models}
\end{figure*}

\section{Reduced BZ models place the two reactions in distinct dynamical scaling classes} \label{sec:models}

We now apply the framework of Section \ref{sec:theory} to two reduced relaxation oscillator models of the BZ reaction. For the classical (catalyzed) reaction, we will see that temperature enters primarily through the nullclines: this places it in Case 2, where a canard explosion hides the Hopf bifurcation and the period scales Arrhenius-like across the entire experimental range. For the uncatalyzed reaction, temperature enters primarily through the timescale separation: this places it in Case 1, where the period scales biphasically as the system loses timescale separation and approaches the Hopf bifurcation. Below, we state the main results but refer to the self-contained Sections IV and V of the Supplemental Material for detailed calculations.\\

\textbf{The classical BZ reaction: temperature-dependent nullclines.}
We start from the two-variable Oregonator, a reduced model of the classical BZ reaction derived from the Field-Körös-Noyes (FKN) mechanism \cite{field1972oscillations, tyson1980target}:
\begin{align} \label{eq:oreg_dyn_eq}
    du/d\tau &= \frac{1}{\epsilon} \left( u(1 - u) - fv \, \frac{u - q}{q + u} \right), \nonumber \\
    dv/d\tau &= u - v.
\end{align}
Here $u$ and $v$ are the rescaled \ce{HBrO2} and \ce{Ce^4+} concentrations, while the \ce{Br-} concentration has been eliminated via a quasi-steady-state approximation \cite{tyson1980target}. The model is schematized in Fig.~\ref{fig:models}A. It has three explicit parameters: $\epsilon \ll 1$ sets the timescale separation, $f$ is a stoichiometric parameter that controls the number of \ce{Br-} ions produced per \ce{Ce^4+} ion in the oxidation of malonic acid, and $q \ll 1$ is a coarse-grained parameter that sets, inter alia, the threshold for the onset of autocatalytic \ce{HBrO2} production. The dynamical time $\tau$ is related to physical time $t$ by a constant rescaling factor $(t/\tau)$. The typical oscillation profile (Fig.~\ref{fig:models}B) shows the characteristic relaxation signature: slow excursions alternated with rapid switches of $u$.

In Section IV of the Supplemental Material we show, based on the analysis in \cite{tyson1980target}, that $q$, $\epsilon$, and $(t/\tau)$ all arise as products of elementary rate constants, so each is naturally assigned an Arrhenius temperature dependence. Later in Section~\ref{sec:data_analysis}, we will show experimentally that the temperature dependence of $\epsilon$ is negligible compared to that of $q$. For now, we therefore take $\epsilon$ constant and write
\begin{align} \label{eq:temp_q}
    q = q_0 \exp\!\left(\frac{E_{q}}{RT}\right), \qquad (t/\tau) = t_0 \exp\!\left(\frac{E_t}{RT}\right).
\end{align}
A similar model has been explored by Pullela \textit{et al.} \cite{pullela_temperature_2009}. With increasing temperature, $q$ grows exponentially, shifting the $u$-nullcline (Fig.~\ref{fig:models}C). For sufficiently large $q$, the limit cycle is destroyed at a Hopf bifurcation (Fig.~\ref{fig:models}D). Depending on $f$ and $q$ this bifurcation can be supercritical or subcritical, but a subcritical bifurcation requires $f<1$, outside the experimentally relevant range $f\in(1,2)$. We therefore fix $f=1$ and treat the bifurcation as supercritical throughout.

This is the Case 2 scenario of Section~\ref{sec:theory}: temperature enters only through the nullclines, while $\epsilon$ remains small and approximately constant. The theory then predicts a single Arrhenius scaling of the period,
\begin{align} \label{eq:BZperiodArr}
    \log P = \log P_0 + \frac{E_t}{RT},
\end{align}
across the entire experimentally accessible temperature range, with the Hopf limit hidden by a canard explosion. A more detailed calculation reveals that $P_0$ is not strictly constant but varies only logarithmically with temperature (see Supplemental Material, Section IV). Fig.~\ref{fig:models}E confirms this: the simulated period follows the RO limit over essentially the full range, with the transition to the Hopf limit compressed into a temperature window too narrow to resolve experimentally.\\

\textbf{The uncatalyzed BZ reaction: temperature-dependent timescale separation.}
For the uncatalyzed reaction, we start from a two-variable skeleton model that captures its essential dynamics \cite{szalai1999}:
\begin{align} \label{eq:szalai_model}
    dx/d\tau &= \gamma \sqrt{x}\,z - \frac{\vartheta x - 1}{\vartheta x + 1}, \nonumber \\
    dz/d\tau &= f - \gamma \sqrt{x}\,z.
\end{align}
Here $x$ and $z$ are the rescaled \ce{HBrO2} and \ce{H2Q} concentrations (with \ce{Br-} eliminated via a quasi-steady-state approximation), as schematized in Fig.~\ref{fig:models}A'. The three free parameters (which are shown to arise as the product of individual rate constants in Section V of the Supplemental Material) are: $f$, the ratio of synthesis rates of \ce{H2Q} and \ce{Br-}; $\gamma$, the autocatalysis rate; and $\vartheta$, the threshold for the onset of autocatalytic \ce{HBrO2} production. The model produces relaxation oscillations (Fig.~\ref{fig:models}B'), although this is not immediately apparent from the equations.

To make the timescale structure explicit, we apply the linear transformation $x \to u = \gamma\vartheta^{-1/2}x$, $z \to v = \gamma\vartheta^{-1/2}(z+x)$, $t \to \tau\gamma\vartheta^{-1/2}t$, and define a timescale separation parameter $\epsilon = \gamma\vartheta^{-3/2}$. The system becomes
\begin{align} \label{eq:UBO_dyn_eq}
    du/d\tau &= \frac{1}{\epsilon} \left( \sqrt{u}(v - u) - \epsilon \, \frac{u-\epsilon^2}{u+\epsilon^2} \right), \nonumber \\
    dv/d\tau &= f - \frac{u-\epsilon^2}{u+\epsilon^2}.
\end{align}
In this form, the model has only two relevant parameters, $f$ and $\epsilon$. Since $f$ and $(u-\epsilon^2)/(u+\epsilon^2)$ are both constrained to lie between 0 and 1 (shown below), the system is a relaxation oscillator with fast variable $u$ and slow variable $v$ for small $\epsilon$. Notably, $\epsilon$ controls both the timescale separation \textit{and} the shape of the $u$-nullcline (Fig.~\ref{fig:models}C').

Sustained oscillations require $f$ to be tuned so that the $v$-nullcline lies between the local maxima of the $u$-nullcline, and $\epsilon$ to be small enough that the $u$-nullcline retains those extrema. The boundary of the oscillatory region in parameter space is a supercritical Hopf bifurcation, which can be determined analytically (Fig.~\ref{fig:models}D'; Supplemental Material, Section V).
Outside this region the system relaxes to a fixed point. 

We will experimentally show in Section~\ref{sec:data_analysis} that the temperature dependence of $f$ is negligible compared to that of $\epsilon$. We therefore take $f$ constant and
\begin{align} \label{eq:temp_epsilon}
    \epsilon = \epsilon_0 \exp\!\left(\frac{E_\epsilon}{RT}\right), \qquad (t/\tau) = t_0 \exp\!\left(\frac{E_t}{RT}\right).
\end{align}
This is a mix of Case 1 and Case 2: temperature affects the timescale separation \textit{and} shifts the nullclines, since $\epsilon$ enters both. However, the simulated period retains the biphasic Arrhenius signature of Case 1 (Fig.~\ref{fig:models}E'). Because $\epsilon$ now controls more than just the timescale separation, the activation energy in the Hopf limit differs from that derived for the generic case in Section~\ref{sec:theory}. The leading contribution of $\epsilon$ to the Hopf period is (see Section V of the Supplemental Material)
\begin{align} \label{eq:Hopf_period_UBZ}
    P_\tau = P_h = P_{h0}(f)\, \epsilon + \mathcal{O}(\epsilon^{3/2}),
\end{align}
with $P_{h0}(f) = 2^{3/2}\pi / [(1+f)^{1/4}(1-f)^{3/4}]$. In the RO limit,
\begin{align} \label{eq:ro_period_UBZ}
    P_\tau = P_{ro} = \frac{\Delta v}{f(1-f)},
\end{align}
where $\Delta v$ is the amplitude of the $v$ oscillations. The amplitude is approximately constant in the RO limit but decreases as the system approaches the Hopf bifurcation; to leading order in $\epsilon$:
\begin{align} \label{eq:ampl_UBZ}
    \Delta v \approx 0.30 - 1.31 \epsilon^{2/3} + o(\epsilon^{2/3}).
\end{align}

The two reduced BZ models thus correspond to the two scenarios found in the prototypical FHN system (Section~\ref{sec:theory}). The classical BZ reaction is a Case 2 oscillator: its period scales Arrhenius-like across the entire experimental range, with no visible deviation as the system approaches the Hopf bifurcation. The uncatalyzed BZ reaction is a Case 1 oscillator (with a Case 2 correction): its period scales biphasically, its amplitude decreases near the Hopf bifurcation, and the waveform should evolve from sawtooth-like at low temperatures to sinusoidal at high temperatures. In the next section, we test these predictions against the experimental data using the temperature dependence of the waveform shape as a proxy for the timescale separation.

\section{Waveform asymmetry predicts three experimental observables} \label{sec:data_analysis}
\begin{figure*}[t]
    \centering
    \includegraphics[width=\linewidth]{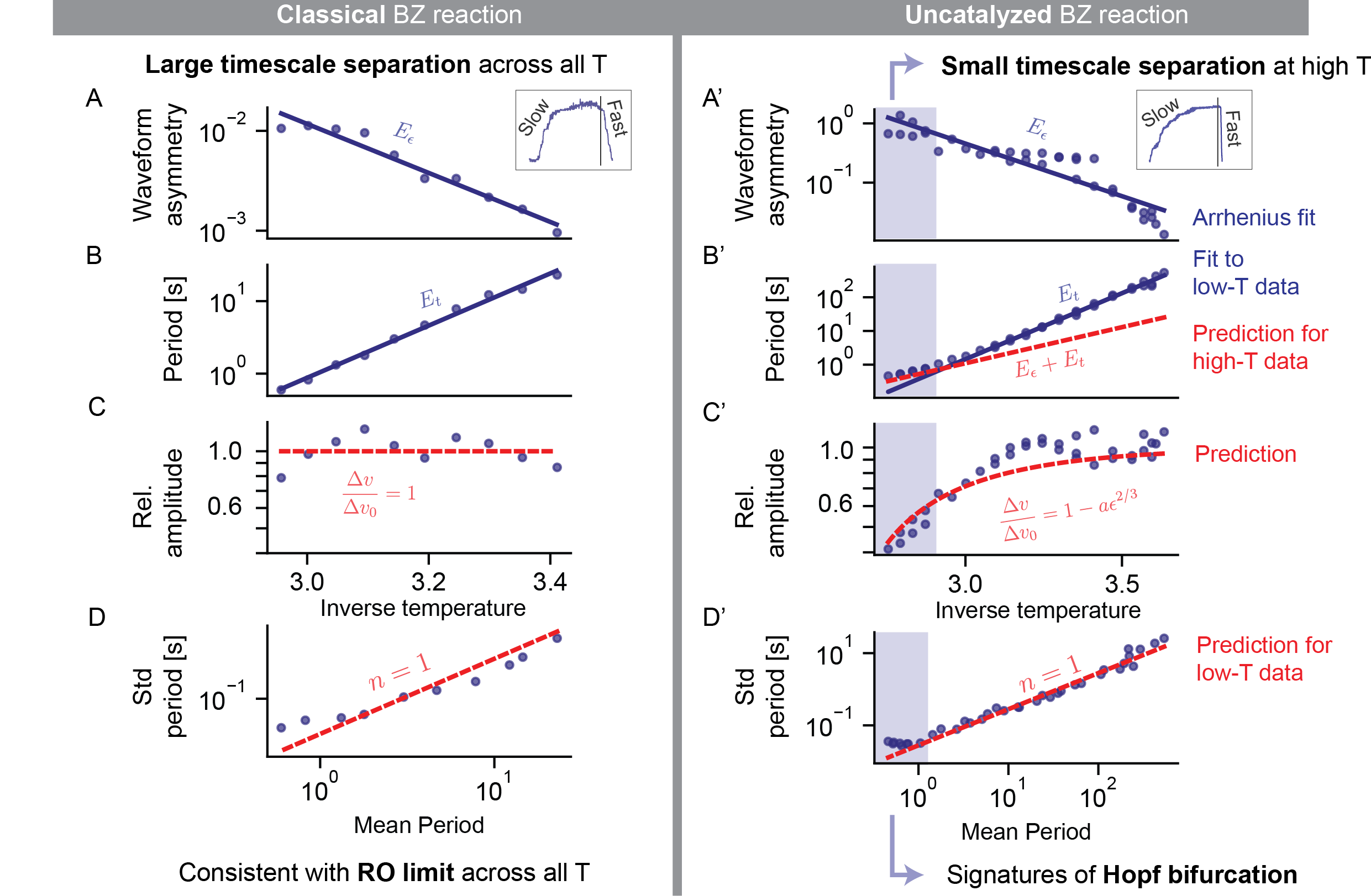}
    \caption{{\textbf{Experimental tests of the theoretical predictions.}} The procedure for respectively the extraction of the observables, the definition of the steady state observables and the temperature fitting is described in Section VII A,B and C of the Supplemental Material.  A/A': (Prediction 1) The waveform asymmetry $R_\epsilon$ in the steady state for the (un)catalyzed BZ reaction in an Arrhenius diagram, together with an Arrhenius fit. For both reactions we defined $R_\epsilon$ as the ratio between fast falling and slow rising phases (see insets). For a perfect sinus wave $R_\epsilon = 1$.
    B/B': (Prediction 2) Period in the steady state, together with an Arrhenius fit at cold temperatures and a theoretical prediction at hot temperatures. 
    C/C': (Prediction 3) Relative amplitude (see text for definition) together with a theoretical prediction. 
    D/D': (Prediction 4) Standard deviation across the steady state as a function of the mean period of the steady state. The theoretical prediction is that of a linear relationship in the relaxation-oscillator regime: $\log \sigma_P = n \log \langle P \rangle + ct.$ with $n = 1$.  }
    \label{fig:steady_state_exp}
\end{figure*}

The models introduced in Section \ref{sec:models} predict that, in the relevant regimes, the temperature dependence of the classical and uncatalyzed BZ reactions arises from the Arrhenius scaling of two parameters. The ratio $(t/\tau)$, with activation energy $E_t$, controls the RO limit, whereas $\epsilon$, with activation energy $E_\epsilon$, controls the Hopf limit.
Both reactions exhibit the same $(t/\tau)$ structure at low temperatures, making $\epsilon$ the key parameter that distinguishes them. In this section, we show that $E_\epsilon$ and the relative magnitude of the timescale separation can be extracted from a single experimental observable: the asymmetry between the rising and falling phases of the waveform. Together, these metrics qualitatively and quantitatively predict the high-temperature scaling of three additional independent observables: the period, the amplitude, and the variance of the period.

For both reactions, oscillating time series data were recorded over a range of temperatures (Fig.~\ref{fig:introduction}A). For each cycle, we extracted the falling-to-rising phase ratio $R_\epsilon$, the period, and the amplitude. A steady state was defined as a sequence of at least five cycles in which amplitude and period varied the least; within each steady state, the three observables were averaged, and the standard deviation of the period was computed. Alternative definitions of the steady state gave consistent results. Details on this analysis are given in Section VII of the Supplemental Material.

Below, we test five specific theoretical predictions using experimental recordings.

\textbf{Prediction 1: Waveform asymmetry follows an Arrhenius law with slope $E_\epsilon$.}
Relaxation oscillations consist of alternating fast and slow phases within each cycle, whereas oscillations near a Hopf bifurcation are smooth and approximately sinusoidal. We quantify this distinction using the asymmetry ratio $R_\epsilon$, defined as the ratio of the durations of the falling and rising phases (See insets in Fig.~\ref{fig:steady_state_exp}A,A'). The ratio is normalized such that $R_\epsilon = 1$ for a perfect sinusoid and scales as $R_\epsilon \propto \epsilon$ in the ODE models (For more details, see Section VIII A of the Supplemental Material).

If $\epsilon$ follows the Arrhenius law, then $R_\epsilon$ should vary linearly in the Arrhenius plot, with a slope determined by $E_\epsilon$. The data confirm this prediction for both reactions. In the classical BZ reaction, $R_\epsilon$ follows an Arrhenius law with $R_{\epsilon}\ll1$ across the entire measured temperature range (Fig.~\ref{fig:steady_state_exp}A), with a slope corresponding to $E_\epsilon = -29$~kJ/mol. This is consistent with the assumption made in Section~\ref{sec:theory} that the classical system remains deep in the RO limit and undergoes a canard-mediated transition that is challenging to resolve experimentally. In the uncatalyzed BZ reaction, $R_\epsilon$ likewise follows the Arrhenius law, but approaches $R_{\epsilon}=1$ at the highest measured temperatures (Fig.~\ref{fig:steady_state_exp}A'), giving $E_\epsilon = -34$~kJ/mol. This indicates that the oscillations become nearly sinusoidal as the system approaches the Hopf bifurcation. In both cases $E_\epsilon < 0$, so that $\epsilon$ grows with temperature and the system is driven toward the Hopf bifurcation, as assumed in Section~\ref{sec:theory}. These values of $E_\epsilon$ are used without further fitting in the predictions below.\\

\textbf{Prediction 2: The effective activation energy of the period shifts from $E_t$ in the RO limit to $E_t + E_\epsilon$ near the Hopf bifurcation.}
Combined with the model results of Section~\ref{sec:models}, the measured value of $R_\epsilon$ determines whether the period exhibits one Arrhenius regime or two. In the classical BZ reaction, $R_\epsilon$ remains sufficiently small, indicating that the Hopf limit is not reached experimentally. The period is therefore expected to follow a single Arrhenius law over the entire temperature range, with activation energy $E_t$.
In the uncatalyzed BZ reaction, by contrast, the period is expected to exhibit biphasic Arrhenius scaling. At low temperatures, where the RO limit is still valid, its activation energy should be $E_t$; at high temperatures, where $R_\epsilon$ approaches 1, it should shift to $E_t + E_\epsilon$. The difference between the low- and high-temperature Arrhenius slopes of the period should therefore equal the slope of $R_\epsilon$, with $E_\epsilon$ taken independently from the fit in Prediction~1.

The data again confirm these predictions (Fig.~\ref{fig:steady_state_exp}B,B'). For the classical BZ reaction, the period follows a single Arrhenius line across the full temperature range, yielding $E_t = 68.8 \pm 0.2$~kJ/mol. This is consistent with the values of around 70~kJ/mol reported in the BZ literature \cite{koros1974monomolecular, misra1992effect, nagy1996effect, fujieda1991thermal, oliveira2012dinamica, zhao2011identification, nogueira2008time, bansagi2009high}. For the uncatalyzed BZ reaction, the period follows an Arrhenius law at low temperatures, with activation energy $E_t = 75.1 \pm 0.3$~kJ/mol, but deviates from this scaling at high temperatures, where it approaches a second Arrhenius line with activation energy $E_t + E_\epsilon$. Crucially, $E_\epsilon$ here is fixed by the slope of $R_\epsilon$ (Fig.~\ref{fig:steady_state_exp}A'), so the high-temperature line does not involve additional fitting parameters\footnote{In principle, $E_\epsilon$ could also be estimated by fitting the high-temperature deviation of the period, but the onset of the Hopf regime is difficult to identify from the period alone, resulting in substantial uncertainty in the fitted value. Estimating $E_\epsilon$ from $R_\epsilon$ is therefore more direct.}.\\

\textbf{Prediction 3: The amplitude decreases as the system approaches the Hopf bifurcation, with a slope set by $E_\epsilon$.}
The amplitude of the measured electrode potential is not the same as the amplitude of the rescaled variables $v$ in the models \eqref{eq:oreg_dyn_eq} and \eqref{eq:UBO_dyn_eq}; the two are related by a proportionality factor that itself arises as a product of rate constants and is therefore assumed to follow an Arrhenius law. We fit this proportionality factor at low temperatures, where the amplitudes of the dynamical variables are approximately constant ($\Delta v = \Delta v_0$), and then use it to convert the experimental potential amplitudes into relative amplitudes $\Delta v / \Delta v_0$. Near the Hopf bifurcation, $\Delta v$ should decrease, with leading-order temperature dependence given by~\eqref{eq:ampl_UBZ}:
\begin{align} \label{eq:experimental_amplitude}
   \Delta v / \Delta v_0 = 1 - C e^{2 E_\epsilon / (3 RT)},
\end{align}
where $C$ is the only free parameter, fixed by fitting to the high-temperature data.

The classical BZ amplitude is indeed approximately constant across the entire {measured temperature} range (Fig.~\ref{fig:steady_state_exp}C), as expected for an oscillator that stays in the RO limit. The uncatalyzed BZ amplitude decreases at high temperatures (Fig.~\ref{fig:steady_state_exp}C'), and the predicted slope from~\eqref{eq:experimental_amplitude} captures the trend. The quantitative agreement is less strong at intermediate temperatures; this discrepancy likely reflects the oversimplification introduced either by the two-variable model or by our assumption on the Arrhenius-scaled proportionality factor relating electrode potential to the dynamical variable. However, the qualitative prediction -- that amplitude drops near the Hopf bifurcation with a slope set by the same $E_\epsilon$ -- is robustly confirmed.\\

\textbf{Prediction 4: Period variance scales linearly with mean period in the RO limit and grows near the Hopf bifurcation.}
A 1D stochastic analysis of \eqref{eq:oreg_dyn_eq} and \eqref{eq:UBO_dyn_eq} as Markov jump processes (see Supplemental Material, Section VI) predicts that, in the RO limit,
\begin{align}
    \sigma_P \sim \langle P \rangle \sim e^{E_t / (RT)},
\end{align}
{where $\sigma_P$ and $\langle P \rangle$ are the period's standard deviation and mean, respectively}. On a log-log plot of $\sigma_P$ versus $\langle P \rangle$, this should appear as a straight line with slope $n = 1$. Near a Hopf bifurcation, on the other hand, fluctuations are amplified by critical slowing down, and $\sigma_P$ should grow faster than $\langle P \rangle$.

Both reactions follow the predicted $n = 1$ scaling in the RO limit (Fig.~\ref{fig:steady_state_exp}D,D'). For the classical BZ reaction, the linear relationship holds across the entire temperature range, confirming once again that the system remains in the RO limit. For the uncatalyzed BZ reaction, the data fall on the line at low temperatures but curve upward at high temperatures, with $\sigma_P$ approaching $\langle P \rangle$ in magnitude just before the limit cycle is destroyed — the noise signature of an oscillator approaching a Hopf bifurcation. \\

\begin{figure*}
    \centering
    \includegraphics[width=1\linewidth]{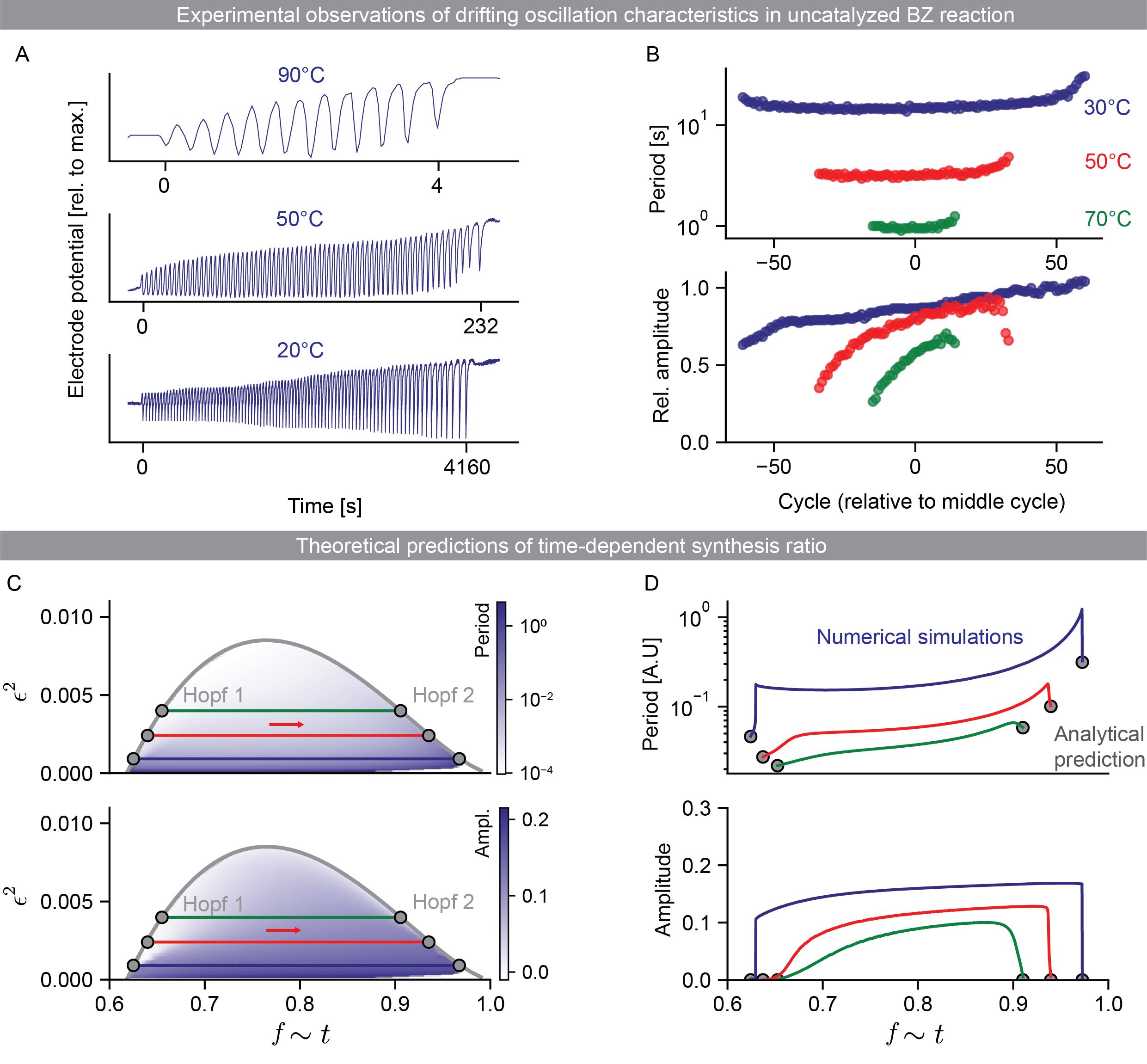}
    \caption{\textbf{Time dependence of oscillations in the uncatalyzed BZ reaction.} A: Experimental electrode potential measurements at three different temperatures. The $y$-axis has been rescaled relative to the maximum and minimum value for each signal.
    B: Peak-to-peak period and relative peak amplitude per cycle (as defined in Section VII A of the Supplemental Material) at different temperatures. For visual reasons, all variables have been plotted on a common x-axis in such a way that the middle cycle overlaps for all temperatures.
    C: The effect of changing $f$ in phase space on the period and the amplitude. Color maps of the steady-state period and the amplitude have been computed numerically.
    D: Simulated steady-state period and amplitude for changing values of $f$ and a fixed value of $\epsilon$. If one assumes that $f$ is proportional to $t$ this can be read as a continuous plot that gives the change of oscillation period and amplitude in function of time. We note that the absolute scale of the $y$-axis is arbitrary and only relative changes are important. The analytical predictions at the supercritical Hopf bifurcations come from \eqref{eq:hopf_period_tau} and the fact that the oscillation amplitude is zero at the Hopf bifurcation (see also Supplemental Material, Section V).}
    \label{fig:transients}
\end{figure*}

\textbf{Prediction 5: Slow changes in the synthesis ratio $f$ drive temporal drift in the uncatalyzed BZ oscillations.}
For fixed parameters, a stable limit cycle has a well-defined period and amplitude. Experimentally, however, these quantities may drift over time as the chemical background evolves. This effect is particularly clear in the uncatalyzed BZ reaction, for which each time series was recorded from the onset to the termination of the oscillatory regime. Representative traces are shown in Fig.~\ref{fig:transients}A, while Fig.~\ref{fig:transients}B shows the cycle-to-cycle evolution of the peak-to-peak period and peak amplitude.

This drift arises because the concentrations of reactants and products change gradually as the reaction proceeds, thereby altering the conditions governing the oscillatory intermediates. In the skeleton model \eqref{eq:UBO_dyn_eq}, we represent this slow evolution by allowing the synthesis ratio $f$ to vary with time (Fig.~\ref{fig:transients}C; Section V of the Supplemental Material). Fig.~\ref{fig:transients}D shows the stationary period and amplitude predicted by the model as functions of $f$ for several temperatures, or equivalently several values of $\epsilon$. Comparison with Fig.~\ref{fig:transients}B shows that a simple linear variation of $f$ with time qualitatively reproduces the observed drift in both period and amplitude.

A time-dependent $f$ also captures the full passage through the oscillatory regime. As the background concentrations evolve, the system transitions from a stationary state to oscillations and then back to a stationary state through two supercritical Hopf bifurcations. At high temperatures, corresponding to large $\epsilon$, these bifurcations lie close together in parameter space (Fig.~\ref{fig:transients}C). The oscillations therefore remain small in amplitude and nearly sinusoidal throughout the oscillatory interval.

At low temperatures, by contrast, the transitions are canard-mediated. Small changes in $f$ away from the Hopf surface then produce an abrupt transition to nonlinear relaxation oscillations. Consequently, the system spends only a short time in the near-Hopf regime as oscillations appear or disappear. When this interval is shorter than a single oscillation period, the near-Hopf dynamics cannot be resolved experimentally. The observed onset and termination of oscillations in Fig.~\ref{fig:transients}A are consistent with these predictions.

\section{Conclusion and discussion} \label{sec:discussion}

Using reduced dynamical models and experimental realizations of the classical and uncatalyzed Belousov-Zhabotinsky (BZ) reactions, we have shown that non-Arrhenius temperature scaling in chemical relaxation oscillators can arise from a temperature-dependent timescale separation between fast and slow variables -- a mechanism that is generic to relaxation oscillators rather than specific to any one circuit.
The classical reaction stays in the relaxation oscillator (RO) limit across the entire {measured temperature} range and shows Arrhenius scaling because a canard explosion hides the Hopf bifurcation. The uncatalyzed reaction crosses smoothly from the RO limit into the Hopf regime at high temperatures, producing the characteristic biphasic curvature on the Arrhenius plot. A key observable to connect theory to experiments was $R_\epsilon$, {the ratio between the oscillation's falling and rising phases, reflecting waveform asymmetry}. By extracting $R_\epsilon$ and its activation energy from the data we qualitatively and quantitatively predicted the behavior of the BZ reaction period, amplitude and phase noise close to the Hopf bifurcation. Below, we discuss what makes this minimal description work, what it reveals about the underlying chemistry, and what it implies for biological oscillators more broadly.\\

\textbf{Why minimal models work.}
The two BZ models we use are minimal in two regards: each compresses a complex reaction mechanism into two coupled ODEs, and each assigns Arrhenius temperature dependence to only a small subset of the resulting parameters. That such a description quantitatively captures eight independent temperature-dependent observables (four per reaction) is initially surprising. We attribute this to two compounding factors: timescale separation and the strong nonlinearity of Arrhenius scaling.

Timescale separation suppresses the fine structure of the reaction network. Fast variables can be eliminated by a quasi-steady-state approximation — which is how the three-variable Oregonator collapses to the two-variable system in Eq.~\eqref{eq:oreg_dyn_eq} (see Supplemental Material, Section IV) — and the resulting reduced dynamics depend on the fast subsystem only through its slow manifold. Whatever happens on the fast timescale is averaged out. This is also why the classical BZ reaction, deep in the RO limit, exhibits clean Arrhenius scaling and near-constant amplitude across the entire experimental range: the Hopf bifurcation predicted by the model is hidden by the canard explosion, and the system is effectively governed by the single rate that sets the flow along the slow manifold. Conversely, the uncatalyzed reaction shows exactly what happens when this separation breaks down: the amplitude drops, the period scales biphasically, and the phase noise grows.

Arrhenius scaling, in turn, amplifies the dominance of the fastest-scaling rate. Reactions with exponential temperature dependence outpace any polynomial form (e.g., Stokes-Einstein diffusion) within a narrow temperature window, and when multiple Arrhenius-scaling rates are present, the one with the largest activation energy dominates the response at temperature extremes. Recent theoretical work has formalized this: Voits and Schwarz \cite{voits_generic_2025} showed that the overall mean-first-passage time of a large network of Arrhenius-scaling transitions reduces, to leading order, again to an Arrhenius-like quadratic-exponential form in inverse temperature. Similarly, in recent work ~\cite{jacobs_beyond_2026}, we showed that cascades of reversible Arrhenius-scaling transitions always lead to Arrhenius scaling at temperature extremes with apparent activation energies determined by specific critical cycles. Together, these results suggest that minimal Arrhenius-parameterized models are not just convenient approximations but also capture the leading-order temperature response of a much larger class of biochemical networks, including the BZ reaction \cite{mundim_temperature_2020} and biological oscillators such as the embryonic cell cycle or circadian rhythms~\cite{ruoff1997modeling,rombouts_mechanistic_2025}. \\

\textbf{What we learn about BZ chemistry.}
Although timescale separation and Arrhenius scaling explain why our minimal models work, they also imply that much of the underlying chemistry is hidden in a few effective parameters. Reducing the multi-step BZ mechanism to a two-variable ODE compresses several elementary rate constants into a handful of model parameters, and the temperature response constrains only certain combinations of the underlying activation energies. Even so, some of these combinations have direct chemical interpretations.

For the classical BZ reaction, the activation energy $E_t$ of the period in the RO limit corresponds to a single elementary step: the consumption of \ce{Ce^4+} in the oxidation of bromomalonic acid (top blue arrow in Fig.~\ref{fig:models}A; see also the Supplemental Material). The activation energy of this reaction is therefore directly given by $E_t = 69$~kJ/mol, consistent with the canonical value of around 70~kJ/mol from prior BZ studies \cite{koros1974monomolecular, misra1992effect, nagy1996effect, fujieda1991thermal, oliveira2012dinamica, zhao2011identification, nogueira2008time, bansagi2009high} and with Körös's original argument that this single reaction is rate-limiting for the classical BZ oscillations \cite{koros1974monomolecular}.

More strikingly, the waveform-derived energy $E_\epsilon$ gives access to a \textit{second} elementary activation energy, that of the autocatalytic production of \ce{HBrO2} (bottom blue arrow in Fig.~\ref{fig:models}A). Combined with $E_t$, this gives a value of 22~kJ/mol for the autocatalytic step (see Supplemental Material, Section IV) — a number not previously extracted from temperature-scaling data of BZ oscillations, accessible here only because the {oscillation waveform} gives an independent constraint on the timescale separation.

For the uncatalyzed reaction, the mapping from model parameters to elementary activation energies is less direct: both $E_t$ and $E_\epsilon$ correspond to linear combinations of four elementary activation energies in the Szalai-Györgyi-Körös mechanism (see Supplemental Material, Section V). Inverting the two measured combinations to recover individual activation energies would require additional measurements. Nonetheless, the temperature dependence of the timescale separation gives access to the relative activation energies between the autocatalytic and substrate-decay branches of the mechanism.

These identifications illustrate a broader principle: information about the underlying reaction mechanism that appears lost at the level of the oscillation period can sometimes be recovered from the detailed shape of the waveform. The waveform-based diagnostic adds a second observable to the standard period-fit, providing access to combinations of rate constants that are otherwise hidden by the timescale-separation reduction. A complementary illustration is provided by the slow drift of oscillation properties within a single experimental run: as background concentrations change, the system effectively sweeps through parameter space, and we observe a canard-explosion entry at low temperatures versus a gentle, near-sinusoidal entry at high temperatures (see Fig. \ref{fig:transients}) — another experimental signature of the two-regime structure beyond the four steady-state observables of Section~\ref{sec:data_analysis}. This connects to a growing literature on using waveform features to identify biochemical mechanisms and evolutionary adaptations in biological oscillators \cite{Gelens2015,de_montaigu_natural_2015,zhang_design_2017,jo_waveforms_2018,tyler_inferring_2021,rombouts_mechanistic_2025,gibo_waveform_2025} and suggests that the BZ reaction — accessible over an exceptionally wide temperature range and with two variants that occupy distinct dynamical regimes — is well-suited as a model system for exploring how oscillator dynamics encode chemical structure.\\

\textbf{Implications for biological oscillators: temperature compensation and phase noise.}
The two-regime structure we identified in the BZ reaction has direct implications for biological oscillators, where temperature compensation — the maintenance of a constant period across temperatures — is a defining feature of circadian rhythms and other biological clocks \cite{pittendrigh1954temperature, hastings1957mechanism, forger_biological_2017}. Our framework is thus particularly relevant for circadian rhythms commonly modeled as relaxation oscillators \cite{Hong2007, fu_temperature_2024}. In its language, temperature compensation requires the activation energies of the rate constants within the oscillator to balance such that $E_t \approx 0$. Our analysis adds a further constraint: even when this balance is achieved, robust temperature compensation requires the oscillator to operate \textit{far} from the Hopf bifurcation. Near the bifurcation, the period scales with $E_\epsilon$ rather than $E_t$, so compensation breaks down generically. And even when $E_\epsilon \approx 0$ as well, operating close to the bifurcation requires fine-tuning, because the canard structure of the bifurcation makes the period extremely sensitive to small parameter changes. This is consistent with recent theoretical work by Fu et al.~\cite{fu_temperature_2024}, who showed that timescale separation is necessary for robust temperature compensation precisely because it allows period-lengthening reactions to balance period-shortening ones — but only far from the Hopf bifurcation. The waveform asymmetry analysis we developed here ($R_\epsilon \ll 1$ versus $R_\epsilon \sim 1$) could be used directly to test whether a temperature-compensated oscillator is operating in the safe regime.

A second implication concerns phase noise. We showed that for the two reactions we consider the variance of the period scales linearly with the mean period in the RO limit but grows nonlinearly near the Hopf bifurcation (Fig.~\ref{fig:steady_state_exp}D'). This means that the stochastic fluctuations of an oscillator's period are themselves a diagnostic of how close the system is to a bifurcation — a more sensitive probe than the period or amplitude alone, both of which vary smoothly through the crossover. In biological systems where the underlying dynamical regime is unclear, measuring how period variance scales with mean period could distinguish a deep-RO-limit oscillator from one approaching loss of stability, even when the deterministic features look similar. Furthermore we remark that the phase noise we studied is chemical in origin: it arises due to the inherent stochasticity of chemical reactions. Studying the effect other temperature-dependent sources of noise in biological oscillators is a promising avenue for future researchers.

More broadly, our results show that the BZ reaction -- long studied as a paradigm of nonlinear chemical dynamics -- can serve as an experimentally tractable model system that captures essential dynamical features of biological relaxation oscillators. By linking a small set of experimental observables to predictions from reduced dynamical models, this approach offers mechanistic insight into the dynamical regimes such oscillators encounter as their endogenous or environmental parameters vary.

\section*{Data availability statement}
The raw data and the Jupyter notebooks to numerically simulate the dynamical systems, analyze the data and reproduce all the plots in this work are openly available in GitLab \cite{gitlab_Jacobs_2026}.


%

\end{document}